\documentclass[ALICE,manyauthors]{cernphprep}

\usepackage{units}
\usepackage{epsfig}
\usepackage{hyperref}
\newcommand{\bq}{\begin{equation}}
\newcommand{\eq}{\end{equation}}

\newcommand{\bqq}{\begin{eqnarray}}
\newcommand{\eqq}{\end{eqnarray}}

\newcommand{\cms}{\sqrt{s}}
\newcommand{\cmsNN}{\sqrt{s_{\rm NN}}}

\newcommand{\etain}[1]{$|\eta|$~$<$~$#1$}

\newcommand{\pt}{\ensuremath{p_{\rm{t}}}}
\newcommand{\pta}{\ensuremath{p_{\rm t, assoc}}}
\newcommand{\ptt}{\ensuremath{p_{\rm t, trig}}}

\newcommand{\Dphi}{\Delta\varphi}
\newcommand{\Deta}{\Delta\eta}
\newcommand{\Ntrig}{N_{\rm trig}}
\newcommand{\Nassoc}{N_{\rm assoc}}
\newcommand{\icp}{I_{\rm CP}}
\newcommand{\iaa}{I_{\rm AA}}

\newcommand{\gmom}    {\mbox{${\rm GeV}/c$}}

\newcommand{\df}     {\mbox{${\rm d}$}}

\newcommand{\figref}[1]{Figure~\ref{#1}}

\newcommand{\bfigFullPage}{\begin{figure} \begin{center} \vspace{0pt}}
\newcommand{\bfig}[1][t!]{\begin{figure*}[#1] \begin{center}}
\newcommand{\efig}{\end{center} \end{figure*}}
\newcommand{\btab}[1][t!]{\begin{table*}[#1] \begin{center}}
\newcommand{\etab}{\end{center} \end{table*}}

\begin{document}%
%
%
\begin{titlepage}
\PHnumber{2011-161}                 
\PHdate{07 December 2011}              
%
%
\title{Particle-yield modification in jet-like azimuthal di-hadron
  correlations in Pb--Pb collisions at $\cmsNN = \unit[2.76]{TeV}$}
\ShortTitle{Particle-yield modification in jet-like azimuthal di-hadron
  correlations in Pb--Pb collisions at $\cmsNN = \unit[2.76]{TeV}$}   
%
\Collaboration{ALICE Collaboration%
         \thanks{See Appendix~\ref{app:collab} for the list of collaboration 
                      members}}
\ShortAuthor{ALICE Collaboration}      
\begin{abstract}
  The yield of charged particles associated with high-$\pt$ trigger
  particles ($8 < \pt < 15 \, \gmom$) is measured with the ALICE detector in Pb--Pb
  collisions at $\cmsNN = \unit[2.76]{TeV}$ relative to
  proton-proton collisions at the same energy.
  The conditional per-trigger yields are extracted from the narrow jet-like correlation peaks
  in azimuthal di-hadron correlations.
  In the 5\% most central collisions, we observe that the
  yield of associated charged particles with transverse momenta $\pt >
  3 \, \gmom$ on the away-side drops to about 60\% of that observed in pp
  collisions, while on the near-side a moderate enhancement of 20-30\% is found.
\end{abstract}
\end{titlepage}
\setcounter{page}{2}

\section{Introduction}

Ultra-relativistic heavy ion-collisions produce
the quark--gluon plasma (QGP), the deconfined state of quarks and
gluons, and are used to explore its properties.  In the last decade, important
information about the dynamical behavior of the QGP has been obtained
from the study of hadron jets, the fragmentation products of high
transverse momentum ($\pt$) partons that are produced in
initial hard scatterings of partons from the incoming nuclei
\cite{enterria,whitepapers}. It is generally accepted that prior to hadronization, partons lose energy in the
high color-density medium due to gluon
radiation and multiple collisions.
These phenomena are broadly known as jet quenching \cite{bjorken}.

The energy loss was first observed at the Relativistic Heavy Ion
Collider (RHIC) in Au--Au collisions at $\cmsNN = \unit[130]{GeV}$ as
a suppression of hadron yields with respect to the reference from pp collisions at high $\pt$ (\unit[3-6]{\gmom})
\cite{raaPHENIX, raaSTAR}.
At RHIC, distributions in relative azimuth
$\Dphi = \varphi_{\rm trig} - \varphi_{\rm assoc}$ between associated particles with
transverse momenta $\pta$ and trigger particles with $\ptt$ have been measured. These studies indicate
that the peak shapes from high-$\pt$ ($\ptt > 4 \, \gmom$ and $2 \, \gmom < \pta < \ptt$) di-hadron correlations in
central Au--Au collisions are similar to  those in small systems like pp and
d--Au~\cite{iaastar0, iaaphenix}, where correlations are dominated by jet fragmentation.
The near-side peak at $\Dphi = 0$ is comparable in magnitude between
all collision systems, while the away-side peak at $\Dphi = \pi$ is
strongly suppressed.
In central Au--Au collisions at $\cmsNN =
\unit[200]{GeV}$, the suppression amounts to a factor of $3-5$ in the
range $0.35 < \pta/\ptt < 0.95$ for $8 < \ptt < 15 \, \gmom$ and $\pta > 3 \, \gmom$ \cite{iaastar}.

At the LHC, the suppression of charged hadrons in central Pb--Pb collisions at $\cmsNN =
\unit[2.76]{TeV}$ increases and the nuclear modification factor $R_{\rm AA}$ drops to 0.14 around
$7 \, \gmom$ \cite{raa}. Furthermore, a strong di-jet energy asymmetry has been reported by the ATLAS and CMS
collaborations \cite{dijetasymmatlas,dijetasymmcms}. A detailed study of the overall momentum
balance in the di-jet events shows evidence for sizable low-$\pt$ radiation outside the cone of the
subleading jet \cite{dijetasymmcms}. These analyses use full event-by-event reconstruction of
di-jets for leading jet transverse momenta above 100$ \, \gmom$.  At lower
transverse momenta ($p_{\rm t,jet} < 50 \, \gmom$) background fluctuations due to the underlying
event dominate \cite{ckb} and event-by-event jet reconstruction becomes difficult.  
Hence, di-hadron
correlations are an interesting alternative probe.
Measurements of di-hadron correlations in central Pb--Pb collisions compared to \textsc{Pythia} 8 \cite{pythia8} pp simulations have been presented in \cite{cms_azimuthalcorrelations}.

The extraction of the particle yield associated with a jet requires
the removal of correlated background primarily of
collective origin (e.g., flow) at lower $\pt$.
This is non-trivial and, therefore, we
concentrate in this letter on a regime
where jet-like correlations dominate over collective effects:
$8 < \ptt < 15 \,\gmom$ for the trigger particle and $\pta > 3 \, \gmom$ for the
associated particle \cite{alice_decomposition}.
We present ratios of yields
of
central to peripheral collisions ($\icp$) and, for different centralities, of Pb--Pb to pp collisions ($\iaa$).
$\iaa$ probes the interplay between the parton production spectrum, the relative importance of 
quark--quark, gluon--gluon and quark--gluon final states, and energy loss in the medium. 
On the near-side, $\iaa$ provides information about the fragmenting jet leaving the medium, while on the away-side it additionally reflects the probability that the recoiling parton survives the passage through the medium.
The sensitivity of $\iaa$ and $R_{\rm AA}$ to different properties of the medium makes the combination particularly effective in constraining jet quenching models \cite{Zhang:2007ja,Armesto:2009zi}.

\section{Detector, Data Sets and Analysis}

The ALICE detector is described in detail in \cite{alice}. The Inner
Tracking System (ITS) and the Time Projection Chamber (TPC) are used
for vertex finding and tracking.
The collision centrality is determined with the forward scintillators (VZERO) as well as for the estimation of the systematic uncertainty with the first two layers of the ITS (Silicon Pixel Detector, SPD) and the Zero Degree Calorimeters (ZDCs).
Details
can be found in \cite{mult_paper}.  The main tracking detector is the TPC which allows reconstruction of good-quality tracks with a pseudorapidity coverage of \etain{1.0} uniform in azimuth. The reconstructed vertex
is used to select primary track candidates and to
constrain the $\pt$ of the track.

In this analysis 14 million minimum-bias Pb--Pb events recorded in fall 2010 at $\cmsNN = \unit[2.76]{TeV}$ as well as 37 million pp events from March 2011 ($\cms = \unit[2.76]{TeV}$) are used. 
These include only events where the TPC was fully efficient to ensure uniform azimuthal acceptance. Events are accepted which have a reconstructed
vertex less than $\unit[7]{cm}$ from the nominal interaction point in beam direction. Tracks are selected by
requiring at least 70 (out of up to 159) associated clusters in the TPC, and a $\chi^2$ per space point
of the momentum fit smaller than 4 (with 2 degrees of freedom per space point). In addition, tracks
are required to originate from within \unit[2.4]{cm} (\unit[3.2]{cm}) in transverse (longitudinal)
distance from the primary vertex.

\begin{figure}
  \centering
    \includegraphics[width=0.5\linewidth]{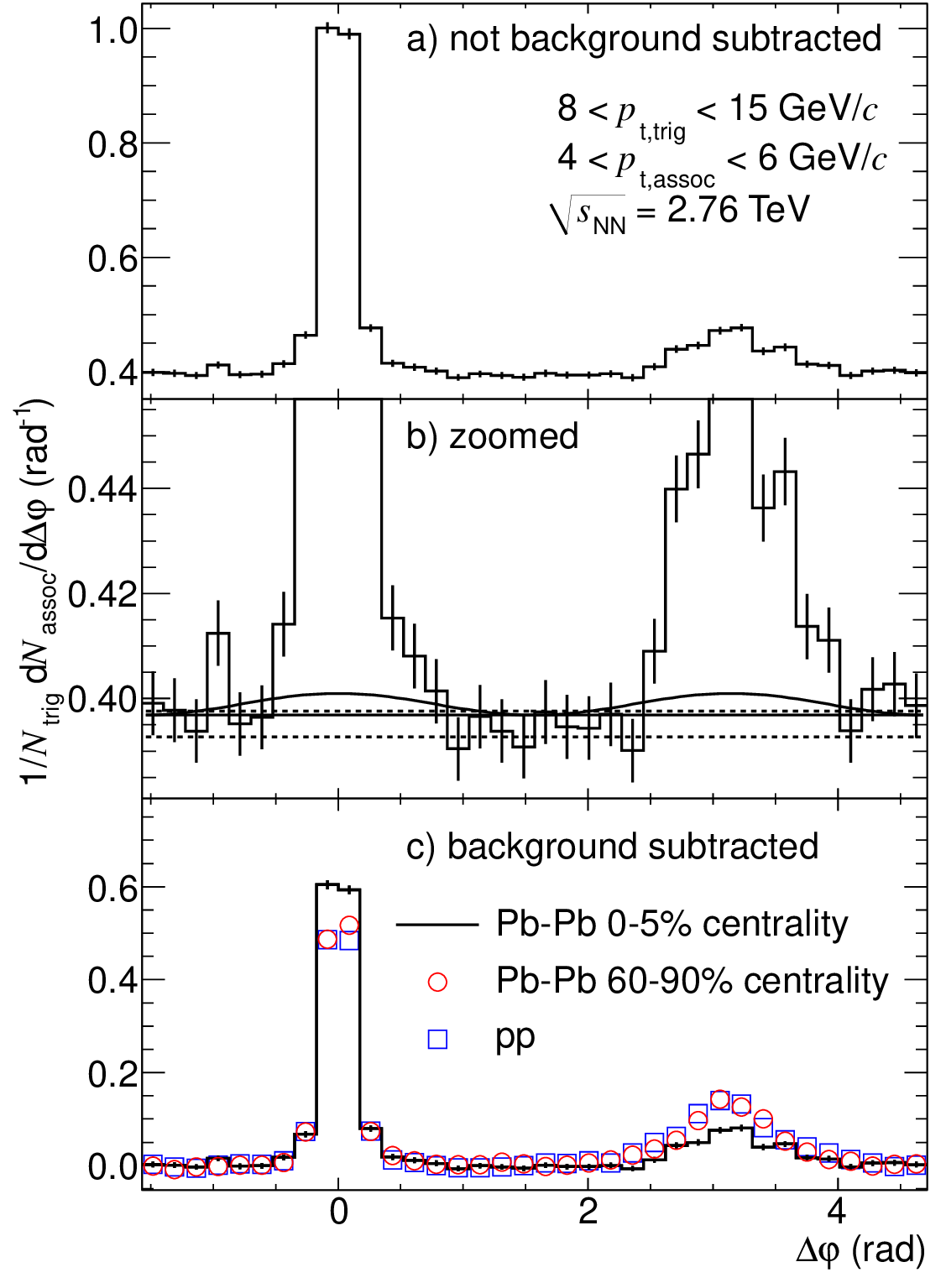}
    \caption{\label{fig_pedestal} Corrected per-trigger pair yield for $4 < \pta < \unit[6]{GeV/\emph{c}}$ for central
    Pb--Pb events (histogram), peripheral Pb--Pb events (red circles) and pp events (blue squares).
    a) azimuthal correlation; b) zoom on
    the region where the pedestal values (horizontal lines) and the $v_2$
    component ($\cos 2 \Dphi$) are indicated. Solid lines are used in the yield extraction while
    the dashed lines are used for the estimation of the uncertainty of the pedestal calculation;
    c) background-subtracted distributions using the flat pedestal.
    Error bars indicate
    statistical uncertainties only.}
\end{figure}

For the measurement of $\iaa$ and $\icp$ the yield of associated particles per trigger particle is
studied as a function of the azimuthal angle difference $\Dphi$. This distribution is given by
$1/\Ntrig \ \df \Nassoc/\df \Dphi$
where $\Ntrig$ is the number of trigger particles and
$\Nassoc$ is the number of associated particles.
We measure this quantity for all pairs of particles where $\pta <
\ptt$ within \etain{1.0} as a function of $\pta$. Pair acceptance corrections have been evaluated
with a mixed-event technique but found to be negligible for the yield ratios due to the constant
acceptance in $\varphi$ and the same detector conditions for the different data sets.

Corrections for detector efficiency (17-18\% depending on collision system, $\pt$ and centrality)
and contamination (4-8\%) by secondary particles from particle--material interactions, $\gamma$ conversions
and weak-decay products of long-lived particles are applied for trigger and associated particles, separately.
Additional secondary particles correlated with the trigger particle are found close to $\Dphi = 0$
in particular due to decays and $\gamma$ conversions. We correct for this contribution (2-4\%).
These corrections are evaluated with the \textsc{Hijing} 1.36 \cite{hijing} Monte Carlo (MC) generator
which was tuned to reproduce the measured multiplicity density \cite{mult_paper} for Pb--Pb and the
\textsc{Pythia} 6 \cite{pythia6} MC with tune Perugia-0 \cite{perugia0} for pp using in both
cases a detector simulation based on \textsc{Geant3} \cite{geant3}. MC simulations underestimate the number of
secondary particles. Therefore, we study the distribution of the distance of closest approach
between tracks and the event vertex.  The tail of this distribution is dominantly populated by
secondary particles and the comparison of data and MC shows that the secondary yield in MC needs to
be increased by about 10\% (depending on $\pt$). An MC study shows that effects of the event
selection and vertex reconstruction are negligible for the extracted
observables. The correction procedure was validated by comparing corrected simulated events with the MC truth.

\figref{fig_pedestal}a shows a typical distribution of the corrected per-trigger
pair yield before background subtraction.
The fact that the $\Dphi$ distribution is flat outside the near- and away-side region gives us confidence that the background can be estimated with
the zero yield at minimum (ZYAM) assumption \cite{zyam}. 
This procedure estimates the pedestal value by fitting the flat region close to the minimum of the $\Dphi$ distribution
($|\Dphi - \pi/2| <  0.4$) with a
constant. The validity
  of the ZYAM assumption has been questioned
in cases where collective effects
dominate \cite{zyam_critisism,zyam_crit2}; however, for the high-$\pt$
  correlations of this analysis, the narrow width and large amplitude
  of the correlated signal compared to the flow modulation drastically
  reduce the ZYAM bias.
Therefore, we define the integrated associated yield as the signal over a flat background.
\figref{fig_pedestal}b illustrates the background determination. Also indicated is a background shape
accounting for elliptic flow $v_2$, the second coefficient of the particle azimuthal distribution measured with respect to the reaction plane.
It is given by $2v_{\rm 2,trig} v_{\rm 2,assoc} \cos 2 \Dphi$ where $v_{\rm 2,trig}$ ($v_{\rm 2,assoc}$) is 
the elliptic flow of the trigger (associated) particles.
The $v_2$
values are taken from an independent measurement \cite{newflow} of $v_2$ up to $\pt = \unit[5]{GeV/\emph{c}}$.
As an upper limit we use the measured $v_2$ for $\pt = \unit[5]{GeV/\emph{c}}$ also for larger $\pt$ where $v_2$ is expected to decrease.
For
the centrality class 60-90\% no $v_2$ measurement is
available, therefore, as an upper limit, $v_2$ is taken from the
40-50\% centrality class. Since $v_2$ decreases from mid-central to peripheral collisions and the flat pedestal assumes $v_2=0$, this includes all reasonable values of $v_2$.

Contributions from $\Deta$-independent correlations (e.g., due to flow
harmonics at all orders) can also be removed on the near-side (where the jet
peak is centered around $\Deta \approx 0$) by calculating
the per-trigger pair yield in the region $|\Deta| < 1$ and subtracting
the contribution from $1 < |\Deta| < 2$ normalized for the acceptance.
This prescription, which we call the $\eta$-gap method, provides a measurement independent of
the flow strength. 

In Fig.~\ref{fig_pedestal}c the flat-pedestal subtracted distributions of central
and peripheral Pb--Pb collisions are compared to that of pp collisions. 
The integral over those
distributions in the region where the signal is significantly above the background, i.e., within
$\Dphi$ of $\pm 0.7$ and $\pi \pm 0.7$ results in the near- and away-side yields per trigger particle ($Y$),
respectively. 
This procedure samples the same fraction of the signal in Pb--Pb and pp collisions, since
in the $\pt$-range used for this study the width of the peaks is similar for both systems.
The yields are used to compute the ratio $\iaa = Y_{\rm Pb-Pb} / Y_{\rm pp}$
where $Y_{\rm Pb-Pb}$ ($Y_{\rm pp}$) is the yield in Pb--Pb (pp)
collisions and the ratio 
$\icp = Y_{\rm 0\text{-}5\%} / Y_{\rm 60\text{-}90\%}$ where $Y_{\rm 0\text{-}5\%}$ ($Y_{\rm 60\text{-}90\%}$)
is the yield in central
(peripheral) Pb--Pb collisions.

\paragraph{Systematic Uncertainties}

\begin{table}
  \centering
  \begin{tabular}{|c|c|c|c|c|}
    \hline
    Uncertainty         & \multicolumn{2}{c|}{$\iaa$}   & \multicolumn{2}{c|}{$\icp$} \\
                & Near-S. & Away-S. & Near-S. & Away-S. \\
    \hline
    Pedestal calculation        & 5\% & 5-20\% & 5\% & 20\% \\ \hline
    Integration window          & 0   & 3\%    & 0   & 3\%  \\ \hline
    Tracking efficiency         & \multicolumn{4}{c|}{4\%}  \\ \hline
    Two-track effects           & \multicolumn{4}{c|}{$<1\%$} \\ \hline
    Corrections                 & \multicolumn{2}{c|}{2\%} & \multicolumn{2}{c|}{1\%} \\ \hline
    Centrality selection        & \multicolumn{2}{c|}{2\%} & \multicolumn{2}{c|}{3\%} \\ \hline \hline
    Total                       & 7\% & 8-21\% & 7\% & 21\% \\ \hline
  \end{tabular}
  \caption{\label{table_syst} Systematic uncertainties evaluated separately for near-side and away-side. Ranges indicate different values for different centrality ranges: the smaller (larger) number is for peripheral (central) events.}
\end{table}

The uncertainty from the
  pedestal determination has been estimated by comparing different
  pedestal evaluation strategies (see Fig.~\ref{fig_pedestal}b).
  The constant-fit
  region has been shifted and an average of the 8 (out of 36) lowest $\Dphi$ points
has been used. The integration window for the near- and away-side has been varied
between $\pm0.5$ and $\pm0.9$. The effect of detector efficiency and track selection has been
studied by systematically varying the track cuts. Track splitting and merging
effects were assessed by studying the tracking performance as a function of the distance of closest
approach of the track pairs in the detector volume. A bias due to the $\pt$ resolution on the
extracted yields was evaluated by folding the detector resolution with the extracted associated
spectrum and found to be negligible. The sensitivity of the corrections
to details of the MC has been studied by varying the particle composition, the material budget and
the MC generator (using \textsc{Ampt} \cite{ampt} for Pb--Pb and \textsc{Phojet} \cite{phojet} for pp).
Uncertainties in the
centrality determination were evaluated by comparing results obtained with the different centrality estimates from the VZERO, the SPD and ZDCs.
Table~\ref{table_syst} lists the size of the
different contributions to the systematic uncertainties for $\iaa$ and $\icp$ as well as their sum in quadrature.

\section{Results}

\bfig
  \includegraphics[width=\linewidth,trim=0 0 0 3]{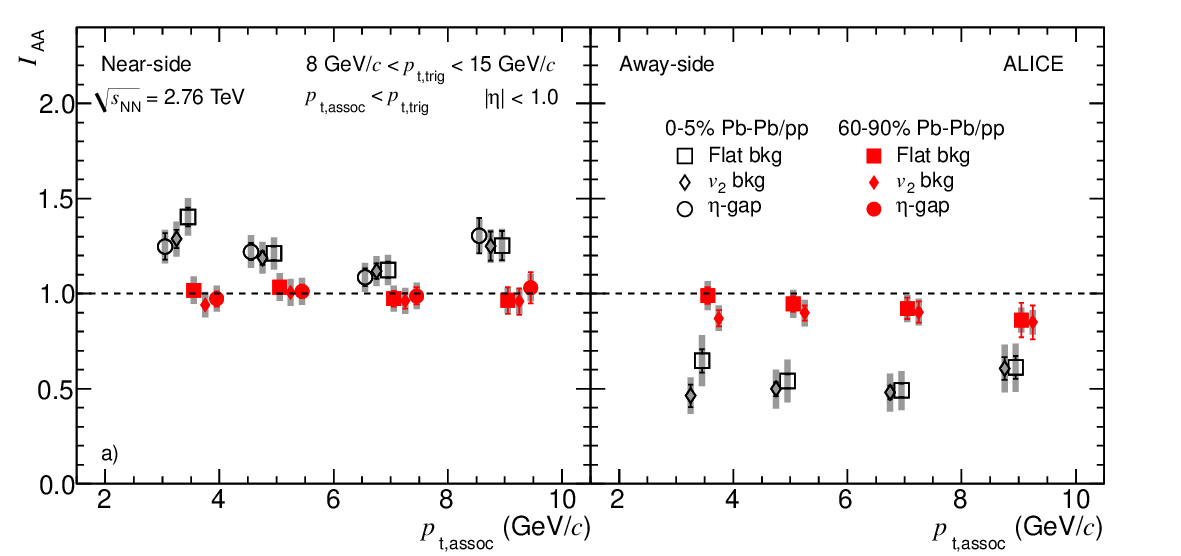}
  \includegraphics[width=\linewidth,trim=0 0 0 3]{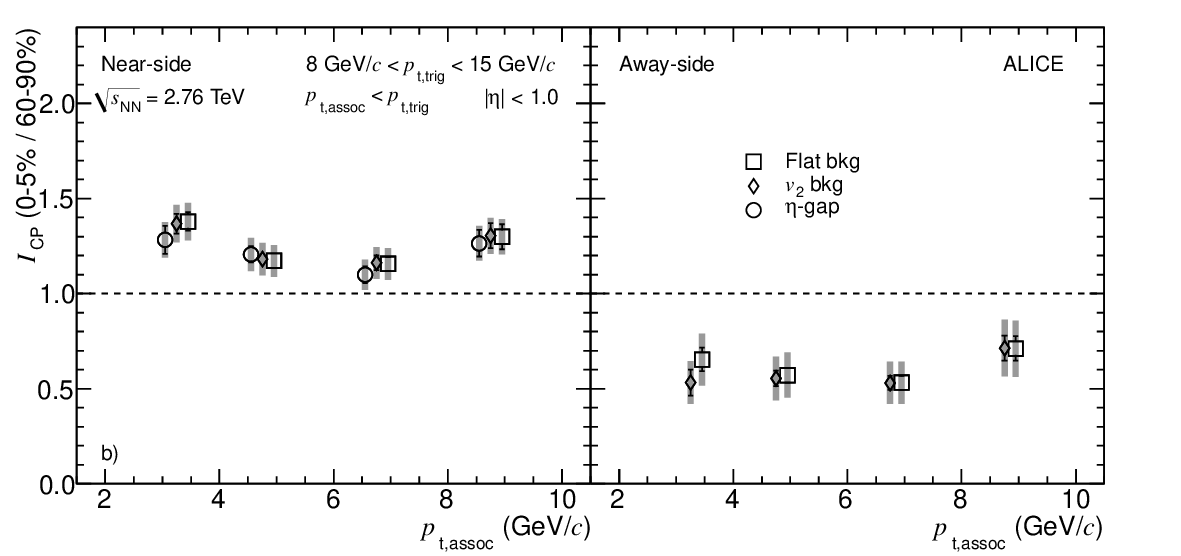}
  \caption{\label{fig_iaa} a) $\iaa$ for central (0-5\% Pb--Pb/pp, open black symbols) and peripheral
    (60-90\% Pb--Pb/pp, filled red symbols) collisions and b) $\icp$. 
    Results using different background subtraction schemes are presented: using a flat pedestal (squares), using $v_2$ subtraction (diamonds) and subtracting the large $|\Deta|$-region (circles, only on the near-side). For details see text. 
For clarity, the data points
    are slightly
    displaced on the $\pta$-axis. The shaded bands denote systematic
    uncertainties.}
  \efig

\figref{fig_iaa}a shows the yield ratio $\iaa$ for central (0-5\% Pb--Pb/pp) and
peripheral (60-90\% Pb--Pb/pp) collisions using the three background subtraction schemes discussed.
 The fact that the only
significant difference between the different background subtraction schemes is in the lowest bin of $\pta$ confirms the
assumption of only a
small bias due to flow anisotropies in this $\pt$ region. 
The influence of higher flow harmonics \cite{newflow} on the background shape can be explicitly estimated: 
including $v_3$, $v_4$ and $v_5$ from \cite{newflow} changes the extracted jet yield by less than 1\%, except for the first bin in $\pta$ in the most central collisions where it is about 8\%. This is consistent with the difference between the data points labeled $v_2$ bkg and $\eta$-gap where the latter includes flow at all orders.
In central collisions, an away-side
suppression ($\iaa \approx 0.6$) is observed which is evidence for in-medium energy loss. Moreover, there is an enhancement above unity of 20-30\% on the near-side which has not been observed with any significance at
lower collision energies at these momenta \cite{iaastar}. 
In peripheral collisions,
both the near- and away-side $\iaa$ measurements approach unity, as expected in the absence of significant medium effects.

\figref{fig_iaa}b shows the yield ratio $\icp$. As
  for $\iaa$, the influence of the flow
modulation is small and only
significant in the lowest $\pta$ bin. $\icp$ is consistent with
$\iaa$ in central collisions with respect to the near-side enhancement
and the away-side suppression.

Comparing this measurement and $R_{\rm AA}$ to models simultaneously will constrain energy-loss mechanisms and model parameters.
Robust conclusions can only be drawn with a systematic comparison of multiple observables with calculations spanning the parameter space and cannot be done with current calculations (e.g. \cite{renknew}). 
Such a study is beyond the scope of this letter.

\paragraph{Comparison to RHIC}
Similar measurements have been performed at RHIC.
Although the same range in $\ptt$ does not necessarily probe the same
parton $\pt$ region at different $\cms$, we assess changes from RHIC to LHC in the following.
The STAR
measurement \cite{iaastar} (which includes only
  statistical uncertainties)
of the near-side $\iaa$ is consistent with unity, albeit with a large uncertainty (18-40\%). On the away-side the result from STAR is about 50\% lower than the results shown in Fig.~\ref{fig_iaa}.
We also calculated $\iaa$ for the 20\% most central events to compare to PHENIX \cite{iaaphenix} (only $v_2$-subtracted data on the away-side available).
For $\pta < \unit[4]{GeV/\emph{c}}$, the flow influence in this centrality interval is about 75\%, too large 
to provide a reliable measurement. For $4 < \pta < \unit[10]{GeV/\emph{c}}$, the $v_2$-subtracted $\iaa$ is $0.5-0.6 \pm 0.08$. This result is slightly larger than results from PHENIX in a similar $\ptt$-region of $7 < \ptt < \unit[9]{GeV/\emph{c}}$: $0.31 \pm 0.07$ and $0.38 \pm 0.11$ for $\pta \approx \unit[3.5]{GeV/\emph{c}}$ and $\unit[5.8]{GeV/\emph{c}}$, respectively.
Based on an analysis in a lower $\pt$-region, where collective effects are significantly larger than in the measurement presented here, the STAR collaboration mentions a slightly enhanced jet-like yield in Au--Au compared to d--Au collisions, but does not assess the effect quantitatively \cite{iaastar2}.
In conclusion, the observed away-side suppression at the LHC is less than at RHIC ($\iaa$ is larger), while the single-hadron suppression $R_{\rm AA}$ is found to be slightly larger ($R_{\rm AA}$ is smaller) than at RHIC \cite{raa}.

\paragraph{Near-Side Enhancement}

These measurements represent the first observation of a significant near-side
enhancement of $\iaa$ and $\icp$ in the $\pt$ region studied. This enhancement suggests
that the near-side parton is also subject to medium effects. 

$\iaa$ is sensitive to (i) a change of the fragmentation function, (ii) a possible change of the quark/gluon jet ratio in the final state due to the different coupling to the medium and (iii) a bias on the parton $\pt$ spectrum after energy loss due to the trigger particle selection.
If the fragmentation function (FF) is softened in the medium, hadrons carry a smaller fraction of the initial parton momentum in Pb--Pb collisions as compared to pp collisions. Therefore, hadrons with a given $\pt$ originate from a larger average parton momentum which may lead to more associated particles and $\iaa > 1$. An increased fraction of gluon (quark) jets has a similar effect than softening (hardening) of the FF and leads to $\iaa > 1$ ($<1$).

A different parton distribution in pp and Pb--Pb collisions can modify $\iaa$ even if fragmentation of a given parton after energy loss is unmodified. 
In particular, in the same transverse momentum region, we see a strong suppression of the trigger particles ($ R_{\rm AA} \approx 0.2$) and the rising slope of $R_{\rm AA}(\pt)$ \cite{raa}.
A similar suppression should apply to partons, leading to a parton distribution after energy loss which is biased towards higher parton $\pt$. 
Therefore, for a fixed trigger $\pt$, the mean parton $\pt$ would be larger in Pb--Pb than in pp, leading to an increase in $\iaa$. This argument can be quantified with the hadron-pair suppression factor $J_{\rm AA}$ \cite{jaa}. $J_{\rm AA}(\ptt,\pta) = R_{\rm AA}(\ptt) \iaa(\ptt,\pta)$ is approximately $R_{\rm AA}(\ptt+\pta)$ in this case, and with a rising $R_{\rm AA}$ leads to $\iaa > 1$.

It is likely that all three effects play a role, and following the above arguments, we note that the combined measurement of $R_{\rm AA}$ and $I_{\rm AA}$ is sensitive to the interplay of energy loss and the change of the fragmentation pattern in the medium.

In summary, the modification of the 
per-trigger yield of associated particles,
$\iaa$ and $\icp$, has
been extracted from di-hadron correlations in Pb--Pb collisions at
$\cmsNN = \unit[2.76]{TeV}$. In central collisions, on the away-side, suppression ($\iaa \approx 0.6$) is observed
as expected from strong in-medium energy
loss. On the near-side, a significant enhancement
($\iaa \approx 1.2$) has been reported for the first time.
Along with the measurement of $R_{\rm AA}$, $\iaa$ provides strong constraints on the quenching mechanism in the hot and dense matter produced.

%
\newenvironment{acknowledgement}{\relax}{\relax}
\begin{acknowledgement}
\section{Acknowledgements}
The ALICE collaboration would like to thank all its engineers and technicians for their invaluable contributions to the construction of the experiment and the CERN accelerator teams for the outstanding performance of the LHC complex.
\\
The ALICE collaboration acknowledges the following funding agencies for their support in building and
running the ALICE detector:
 \\
Calouste Gulbenkian Foundation from Lisbon and Swiss Fonds Kidagan, Armenia;
 \\
Conselho Nacional de Desenvolvimento Cient\'{\i}fico e Tecnol\'{o}gico (CNPq), Financiadora de Estudos e Projetos (FINEP),
Funda\c{c}\~{a}o de Amparo \`{a} Pesquisa do Estado de S\~{a}o Paulo (FAPESP);
 \\
National Natural Science Foundation of China (NSFC), the Chinese Ministry of Education (CMOE)
and the Ministry of Science and Technology of China (MSTC);
 \\
Ministry of Education and Youth of the Czech Republic;
 \\
Danish Natural Science Research Council, the Carlsberg Foundation and the Danish National Research Foundation;
 \\
The European Research Council under the European Community's Seventh Framework Programme;
 \\
Helsinki Institute of Physics and the Academy of Finland;
 \\
French CNRS-IN2P3, the `Region Pays de Loire', `Region Alsace', `Region Auvergne' and CEA, France;
 \\
German BMBF and the Helmholtz Association;
\\
General Secretariat for Research and Technology, Ministry of
Development, Greece;
\\
Hungarian OTKA and National Office for Research and Technology (NKTH);
 \\
Department of Atomic Energy and Department of Science and Technology of the Government of India;
 \\
Istituto Nazionale di Fisica Nucleare (INFN) of Italy;
 \\
MEXT Grant-in-Aid for Specially Promoted Research, Ja\-pan;
 \\
Joint Institute for Nuclear Research, Dubna;
 \\
National Research Foundation of Korea (NRF);
 \\
CONACYT, DGAPA, M\'{e}xico, ALFA-EC and the HELEN Program (High-Energy physics Latin-American--European Network);
 \\
Stichting voor Fundamenteel Onderzoek der Materie (FOM) and the Nederlandse Organisatie voor Wetenschappelijk Onderzoek (NWO), Netherlands;
 \\
Research Council of Norway (NFR);
 \\
Polish Ministry of Science and Higher Education;
 \\
National Authority for Scientific Research - NASR (Autoritatea Na\c{t}ional\u{a} pentru Cercetare \c{S}tiin\c{t}ific\u{a} - ANCS);
 \\
Federal Agency of Science of the Ministry of Education and Science of Russian Federation, International Science and
Technology Center, Russian Academy of Sciences, Russian Federal Agency of Atomic Energy, Russian Federal Agency for Science and Innovations and CERN-INTAS;
 \\
Ministry of Education of Slovakia;
 \\
Department of Science and Technology, South Africa;
 \\
CIEMAT, EELA, Ministerio de Educaci\'{o}n y Ciencia of Spain, Xunta de Galicia (Conseller\'{\i}a de Educaci\'{o}n),
CEA\-DEN, Cubaenerg\'{\i}a, Cuba, and IAEA (International Atomic Energy Agency);
 \\
Swedish Reseach Council (VR) and Knut $\&$ Alice Wallenberg Foundation (KAW);
 \\
Ukraine Ministry of Education and Science;
 \\
United Kingdom Science and Technology Facilities Council (STFC);
 \\
The United States Department of Energy, the United States National
Science Foundation, the State of Texas, and the State of Ohio.
\end{acknowledgement}

\newpage
%
%
\appendix
\section{The ALICE Collaboration}
\label{app:collab}
%
\begingroup
\small
\begin{flushleft}
K.~Aamodt\Irefn{0}\And
B.~Abelev\Irefn{1}\And
A.~Abrahantes~Quintana\Irefn{2}\And
D.~Adamov\'{a}\Irefn{3}\And
A.M.~Adare\Irefn{4}\And
M.M.~Aggarwal\Irefn{5}\And
G.~Aglieri~Rinella\Irefn{6}\And
A.G.~Agocs\Irefn{7}\And
A.~Agostinelli\Irefn{8}\And
S.~Aguilar~Salazar\Irefn{9}
% ...
\renewcommand\labelenumi{\textsuperscript{\theenumi}~}
\section*{Affiliation notes}
\renewcommand\theenumi{\roman{enumi}}
\begin{Authlist}
\item \Adef{0}Also at Laboratoire de Physique Corpusculaire (LPC), Clermont Universit\'{e}, Universit\'{e} Blaise Pascal, CNRS--IN2P3, Clermont-Ferrand, France
\item \Adef{1}Also at Sezione INFN, Bologna, Italy
\item \Adef{2}Now at Physikalisches Institut, Ruprecht-Karls-Universit\"{a}t Heidelberg, Heidelberg, Germany
\item \Adef{3}Now at Laboratoire de Physique Corpusculaire (LPC), Clermont Universit\'{e}, Universit\'{e} Blaise Pascal, CNRS--IN2P3, Clermont-Ferrand, France
\item \Adef{4}Also at  Dipartimento di Fisica dell'Universita, Udine, Italy 
\item \Adef{5}Now at Sezione INFN, Turin, Italy
\item \Adef{6}Also at Benem\'{e}rita Universidad Aut\'{o}noma de Puebla, Puebla, Mexico
\item \Adef{7}Also at European Organization for Nuclear Research (CERN), Geneva, Switzerland
\item \Adef{8}Now at European Organization for Nuclear Research (CERN), Geneva, Switzerland
\item \Adef{9}Now at Lawrence Berkeley National Laboratory, Berkeley, California, United States
\item \Adef{10}Now at Centro Fermi -- Centro Studi e Ricerche e Museo Storico della Fisica ``Enrico Fermi'', Rome, Italy
\item \Adef{11}Now at Warsaw University of Technology, Warsaw, Poland
\item \Adef{12}Also at Fachhochschule K\"{o}ln, K\"{o}ln, Germany
\item \Adef{13}Also at Institute of Experimental Physics, Slovak Academy of Sciences, Ko\v{s}ice, Slovakia
\item \Adef{14}Also at  M.V.Lomonosov Moscow State University, D.V.Skobeltsyn Institute of Nuclear Physics, Moscow, Russia 
\item \Adef{15}Also at Laboratoire de Physique Subatomique et de Cosmologie (LPSC), Universit\'{e} Joseph Fourier, CNRS-IN2P3, Institut Polytechnique de Grenoble, Grenoble, France
\item \Adef{16}Now at Dipartimento Interateneo di Fisica `M.~Merlin' and Sezione INFN, Bari, Italy
\item \Adef{17}Now at SUBATECH, Ecole des Mines de Nantes, Universit\'{e} de Nantes, CNRS-IN2P3, Nantes, France
\item \Adef{18}Also at  "Vin\v{c}a" Institute of Nuclear Sciences, Belgrade, Serbia 
\item \Adef{19}Also at Instituto de Ciencias Nucleares, Universidad Nacional Aut\'{o}noma de M\'{e}xico, Mexico City, Mexico
\item \Adef{20}Now at Wayne State University, Detroit, Michigan, United States
\item \Adef{21}Now at Department of Physics, University of Oslo, Oslo, Norway
\item \Adef{22}Now at Eberhard Karls Universit\"{a}t T\"{u}bingen, T\"{u}bingen, Germany
\item \Adef{23}Now at Research Division and ExtreMe Matter Institute EMMI, GSI Helmholtzzentrum f\"ur Schwerionenforschung, Darmstadt, Germany
\item \Adef{24}Now at Institut Pluridisciplinaire Hubert Curien (IPHC), Universit\'{e} de Strasbourg, CNRS-IN2P3, Strasbourg, France
\item \Adef{25}Also at Centro de Investigaciones Energ\'{e}ticas Medioambientales y Tecnol\'{o}gicas (CIEMAT), Madrid, Spain
\item \Adef{26}Now at University of Tokyo, Tokyo, Japan
\item \Adef{27}Also at Nuclear Physics Institute, Academy of Sciences of the Czech Republic, \v{R}e\v{z} u Prahy, Czech Republic
\item \Adef{28}Also at Hua-Zhong Normal University, Wuhan, China
\item \Adef{29}Now at Sezione INFN, Bologna, Italy
\item \Adef{30}Also at Centro Fermi -- Centro Studi e Ricerche e Museo Storico della Fisica ``Enrico Fermi'', Rome, Italy
\end{Authlist}
\section*{Collaboration Institutes}
\renewcommand\theenumi{\arabic{enumi}~}
\begin{Authlist}
\item \Idef{0}Department of Physics and Technology, University of Bergen, Bergen, Norway
\item \Idef{1}Lawrence Livermore National Laboratory, Livermore, California, United States
\item \Idef{2}Centro de Aplicaciones Tecnol\'{o}gicas y Desarrollo Nuclear (CEADEN), Havana, Cuba
\item \Idef{3}Nuclear Physics Institute, Academy of Sciences of the Czech Republic, \v{R}e\v{z} u Prahy, Czech Republic
\item \Idef{4}Yale University, New Haven, Connecticut, United States
\item \Idef{5}Physics Department, Panjab University, Chandigarh, India
\item \Idef{6}European Organization for Nuclear Research (CERN), Geneva, Switzerland
\item \Idef{7}KFKI Research Institute for Particle and Nuclear Physics, Hungarian Academy of Sciences, Budapest, Hungary
\item \Idef{8}Dipartimento di Fisica dell'Universit\`{a} and Sezione INFN, Bologna, Italy
\item \Idef{9}Instituto de F\'{\i}sica, Universidad Nacional Aut\'{o}noma de M\'{e}xico, Mexico City, Mexico
\item \Idef{10}Variable Energy Cyclotron Centre, Kolkata, India
\item \Idef{11}Department of Physics Aligarh Muslim University, Aligarh, India
\item \Idef{12}Gangneung-Wonju National University, Gangneung, South Korea
\item \Idef{13}Institute for Theoretical and Experimental Physics, Moscow, Russia
\item \Idef{14}Russian Research Centre Kurchatov Institute, Moscow, Russia
\item \Idef{15}Sezione INFN, Turin, Italy
\item \Idef{16}Centro Fermi -- Centro Studi e Ricerche e Museo Storico della Fisica ``Enrico Fermi'', Rome, Italy
\item \Idef{17}Bogolyubov Institute for Theoretical Physics, Kiev, Ukraine
\item \Idef{18}Faculty of Engineering, Bergen University College, Bergen, Norway
\item \Idef{19}Frankfurt Institute for Advanced Studies, Johann Wolfgang Goethe-Universit\"{a}t Frankfurt, Frankfurt, Germany
\item \Idef{20}Dipartimento Interateneo di Fisica `M.~Merlin' and Sezione INFN, Bari, Italy
\item \Idef{21}V.~Fock Institute for Physics, St. Petersburg State University, St. Petersburg, Russia
\item \Idef{22}National Institute for Physics and Nuclear Engineering, Bucharest, Romania
\item \Idef{23}Research Division and ExtreMe Matter Institute EMMI, GSI Helmholtzzentrum f\"ur Schwerionenforschung, Darmstadt, Germany
\item \Idef{24}Institut f\"{u}r Kernphysik, Westf\"{a}lische Wilhelms-Universit\"{a}t M\"{u}nster, M\"{u}nster, Germany
\item \Idef{25}Rudjer Bo\v{s}kovi\'{c} Institute, Zagreb, Croatia
\item \Idef{26}Sezione INFN, Padova, Italy
\item \Idef{27}Sezione INFN, Bologna, Italy
\item \Idef{28}SUBATECH, Ecole des Mines de Nantes, Universit\'{e} de Nantes, CNRS-IN2P3, Nantes, France
\item \Idef{29}Institut f\"{u}r Kernphysik, Johann Wolfgang Goethe-Universit\"{a}t Frankfurt, Frankfurt, Germany
\item \Idef{30}Laboratoire de Physique Subatomique et de Cosmologie (LPSC), Universit\'{e} Joseph Fourier, CNRS-IN2P3, Institut Polytechnique de Grenoble, Grenoble, France
\item \Idef{31}Departamento de F\'{\i}sica de Part\'{\i}culas and IGFAE, Universidad de Santiago de Compostela, Santiago de Compostela, Spain
\item \Idef{32}Oak Ridge National Laboratory, Oak Ridge, Tennessee, United States
\item \Idef{33}Helsinki Institute of Physics (HIP) and University of Jyv\"{a}skyl\"{a}, Jyv\"{a}skyl\"{a}, Finland
\item \Idef{34}Sezione INFN, Catania, Italy
\item \Idef{35}Commissariat \`{a} l'Energie Atomique, IRFU, Saclay, France
\item \Idef{36}Laboratoire de Physique Corpusculaire (LPC), Clermont Universit\'{e}, Universit\'{e} Blaise Pascal, CNRS--IN2P3, Clermont-Ferrand, France
\item \Idef{37}Institute of Experimental Physics, Slovak Academy of Sciences, Ko\v{s}ice, Slovakia
\item \Idef{38}Institute of Physics, Bhubaneswar, India
\item \Idef{39}Dipartimento di Fisica e Astronomia dell'Universit\`{a} and Sezione INFN, Catania, Italy
\item \Idef{40}School of Physics and Astronomy, University of Birmingham, Birmingham, United Kingdom
\item \Idef{41}The Henryk Niewodniczanski Institute of Nuclear Physics, Polish Academy of Sciences, Cracow, Poland
\item \Idef{42}Joint Institute for Nuclear Research (JINR), Dubna, Russia
\item \Idef{43}Niels Bohr Institute, University of Copenhagen, Copenhagen, Denmark
\item \Idef{44}Institut Pluridisciplinaire Hubert Curien (IPHC), Universit\'{e} de Strasbourg, CNRS-IN2P3, Strasbourg, France
\item \Idef{45}University of Houston, Houston, Texas, United States
\item \Idef{46}Dipartimento di Fisica Sperimentale dell'Universit\`{a} and Sezione INFN, Turin, Italy
\item \Idef{47}Petersburg Nuclear Physics Institute, Gatchina, Russia
\item \Idef{48}Physics Department, University of Jammu, Jammu, India
\item \Idef{49}Laboratori Nazionali di Frascati, INFN, Frascati, Italy
\item \Idef{50}Dipartimento di Fisica dell'Universit\`{a} and Sezione INFN, Padova, Italy
\item \Idef{51}Faculty of Nuclear Sciences and Physical Engineering, Czech Technical University in Prague, Prague, Czech Republic
\item \Idef{52}Nikhef, National Institute for Subatomic Physics, Amsterdam, Netherlands
\item \Idef{53}Centro de Investigaciones Energ\'{e}ticas Medioambientales y Tecnol\'{o}gicas (CIEMAT), Madrid, Spain
\item \Idef{54}Department of Physics, Ohio State University, Columbus, Ohio, United States
\item \Idef{55}Moscow Engineering Physics Institute, Moscow, Russia
\item \Idef{56}Institute for High Energy Physics, Protvino, Russia
\item \Idef{57}Faculty of Science, P.J.~\v{S}af\'{a}rik University, Ko\v{s}ice, Slovakia
\item \Idef{58}Wayne State University, Detroit, Michigan, United States
\item \Idef{59}Saha Institute of Nuclear Physics, Kolkata, India
\item \Idef{60}Kirchhoff-Institut f\"{u}r Physik, Ruprecht-Karls-Universit\"{a}t Heidelberg, Heidelberg, Germany
\item \Idef{61}Institut de Physique Nucl\'{e}aire d'Orsay (IPNO), Universit\'{e} Paris-Sud, CNRS-IN2P3, Orsay, France
\item \Idef{62}Faculty of Mathematics, Physics and Informatics, Comenius University, Bratislava, Slovakia
\item \Idef{63}Russian Federal Nuclear Center (VNIIEF), Sarov, Russia
\item \Idef{64}Physikalisches Institut, Ruprecht-Karls-Universit\"{a}t Heidelberg, Heidelberg, Germany
\item \Idef{65}Physics Department, University of Cape Town, iThemba LABS, Cape Town, South Africa
\item \Idef{66}Hua-Zhong Normal University, Wuhan, China
\item \Idef{67}Secci\'{o}n F\'{\i}sica, Departamento de Ciencias, Pontificia Universidad Cat\'{o}lica del Per\'{u}, Lima, Peru
\item \Idef{68}Dipartimento di Fisica dell'Universit\`{a} and Sezione INFN, Trieste, Italy
\item \Idef{69}Centro de Investigaci\'{o}n y de Estudios Avanzados (CINVESTAV), Mexico City and M\'{e}rida, Mexico
\item \Idef{70}Dipartimento di Fisica dell'Universit\`{a} and Sezione INFN, Cagliari, Italy
\item \Idef{71}Physics Department, Creighton University, Omaha, Nebraska, United States
\item \Idef{72}Universit\'{e} de Lyon, Universit\'{e} Lyon 1, CNRS/IN2P3, IPN-Lyon, Villeurbanne, France
\item \Idef{73}Universidade Estadual de Campinas (UNICAMP), Campinas, Brazil
\item \Idef{74}Nikhef, National Institute for Subatomic Physics and Institute for Subatomic Physics of Utrecht University, Utrecht, Netherlands
\item \Idef{75}Division of Experimental High Energy Physics, University of Lund, Lund, Sweden
\item \Idef{76}University of Tsukuba, Tsukuba, Japan
\item \Idef{77}Pusan National University, Pusan, South Korea
\item \Idef{78}Sezione INFN, Cagliari, Italy
\item \Idef{79}Benem\'{e}rita Universidad Aut\'{o}noma de Puebla, Puebla, Mexico
\item \Idef{80}Dipartimento di Scienze e Tecnologie Avanzate dell'Universit\`{a} del Piemonte Orientale and Gruppo Collegato INFN, Alessandria, Italy
\item \Idef{81}Instituto de Ciencias Nucleares, Universidad Nacional Aut\'{o}noma de M\'{e}xico, Mexico City, Mexico
\item \Idef{82}Institute of Space Sciences (ISS), Bucharest, Romania
\item \Idef{83}Universidade de S\~{a}o Paulo (USP), S\~{a}o Paulo, Brazil
\item \Idef{84}Dipartimento di Fisica `E.R.~Caianiello' dell'Universit\`{a} and Gruppo Collegato INFN, Salerno, Italy
\item \Idef{85}Sezione INFN, Bari, Italy
\item \Idef{86}Soltan Institute for Nuclear Studies, Warsaw, Poland
\item \Idef{87}Sezione INFN, Rome, Italy
\item \Idef{88}Department of Physics, University of Oslo, Oslo, Norway
\item \Idef{89}Lawrence Berkeley National Laboratory, Berkeley, California, United States
\item \Idef{90}Institute for Nuclear Research, Academy of Sciences, Moscow, Russia
\item \Idef{91}Sezione INFN, Trieste, Italy
\item \Idef{92}Physics Department, University of Athens, Athens, Greece
\item \Idef{93}Chicago State University, Chicago, United States
\item \Idef{94}Warsaw University of Technology, Warsaw, Poland
\item \Idef{95}Universidad Aut\'{o}noma de Sinaloa, Culiac\'{a}n, Mexico
\item \Idef{96}Physics Department, University of Rajasthan, Jaipur, India
\item \Idef{97}Technical University of Split FESB, Split, Croatia
\item \Idef{98}Yerevan Physics Institute, Yerevan, Armenia
\item \Idef{99}University of Tokyo, Tokyo, Japan
\item \Idef{100}Department of Physics, Sejong University, Seoul, South Korea
\item \Idef{101}Indian Institute of Technology, Mumbai, India
\item \Idef{102}Institut f\"{u}r Kernphysik, Technische Universit\"{a}t Darmstadt, Darmstadt, Germany
\item \Idef{103}Yonsei University, Seoul, South Korea
\item \Idef{104}Zentrum f\"{u}r Technologietransfer und Telekommunikation (ZTT), Fachhochschule Worms, Worms, Germany
\item \Idef{105}California Polytechnic State University, San Luis Obispo, California, United States
\item \Idef{106}China Institute of Atomic Energy, Beijing, China
\item \Idef{107}Institute of Physics, Academy of Sciences of the Czech Republic, Prague, Czech Republic
\item \Idef{108}The University of Texas at Austin, Physics Department, Austin, TX, United States
\item \Idef{109}University of Tennessee, Knoxville, Tennessee, United States
\item \Idef{110}Dipartimento di Fisica dell'Universit\`{a} `La Sapienza' and Sezione INFN, Rome, Italy
\item \Idef{111}Hiroshima University, Hiroshima, Japan
\item \Idef{112}Budker Institute for Nuclear Physics, Novosibirsk, Russia
\item \Idef{113}Laboratori Nazionali di Legnaro, INFN, Legnaro, Italy
\item \Idef{114}Purdue University, West Lafayette, Indiana, United States
\item \Idef{115}Centre de Calcul de l'IN2P3, Villeurbanne, France 
\end{Authlist}
\endgroup
%
%

\begin{thebibliography}{10}
\bibitem{enterria} D. d'Enterria, Springer Verlag. Landolt-Boernstein Vol. 1-23A, arXiv:0902.2011v2 [nucl-ex].

\bibitem{whitepapers}
  I.~Arsene {\it et al.}  [BRAHMS Collaboration],
  Nucl.\ Phys.\  A {\bf 757}, 1 (2005);
  K.~Adcox {\it et al.}  [PHENIX Collaboration],
  Nucl.\ Phys.\  A {\bf 757}, 184 (2005);
  B.~B.~Back {\it et al.},
  Nucl.\ Phys.\  A {\bf 757}, 28 (2005);
  J.~Adams {\it et al.}  [STAR Collaboration],
  Nucl.\ Phys.\  A {\bf 757}, 102 (2005).

\bibitem{bjorken} J. D. Bjorken, FERMILAB-PUB-82-059-THY (1982); %
%
M.~Gyulassy and M.~Plumer, Phys.\ Lett.\, {\bf B243}, 432 (1990);
%
X.-N.~Wang and M.~Gyulassy, Phys.\ Rev.\ Lett.\, {\bf 68}, 1480 (1992);
%

\bibitem{raaPHENIX} K. Adcox {\it et al.} [PHENIX Collaboration], Phys.\ Rev.\ Lett.\ {\bf 88}, 022301 (2001).

\bibitem{raaSTAR}   C. Adler {\it et al.} [STAR Collaboration],   Phys.\ Rev.\ Lett.\ {\bf 89}, 202301 (2002).

\bibitem{iaastar0}   J.~Adams {\it et al.} [STAR Collaboration],
Phys. Rev. Lett. {\bf 91}, 072304 (2003).

\bibitem{iaaphenix}   A.~Adare {\it et al.} [PHENIX Collaboration],
 Phys.\ Rev.\ Lett.\  {\bf 104 }, 252301 (2010).

\bibitem{iaastar}   J.~Adams {\it et al.} [STAR Collaboration],
  Phys.\ Rev.\ Lett.\  {\bf 97 }, 162301 (2006).

\bibitem{raa}   K.~Aamodt {\it et al.} [ALICE Collaboration],
  Phys.\ Lett.\  {\bf B696 }, 30-39 (2011).

\bibitem{dijetasymmatlas}
  G.~Aad {\it et al.}  [ATLAS Collaboration],
  Phys.\ Rev.\ Lett.\  {\bf 105}, 252303 (2010).

\bibitem{dijetasymmcms}
  S.~Chatrchyan {\it et al.}  [CMS Collaboration],
  Phys.\  Rev.\  {\bf C84}, 024906 (2011).

\bibitem{ckb}
  C.~Klein-Boesing [ALICE Collaboration],
  J. Phys. G: Nucl. Part. Phys. 38 (2011) 124088.

\bibitem{pythia8}
  T.~Sjostrand, S.~Mrenna and P.~Z.~Skands,
  Comput.\ Phys.\ Commun.\  {\bf 178}, 852 (2008).



\bibitem{cms_azimuthalcorrelations}
  S.~Chatrchyan {\it et al.}  [CMS Collaboration],
  JHEP {\bf 07}, 076 (2011).


\bibitem{alice_decomposition}
  K.~Aamodt {\it et al.} [ALICE Collaboration],
  arXiv:1109.2501 [nucl-ex].


\bibitem{Zhang:2007ja}
  H.~Zhang, J.~F.~Owens, E.~Wang and X.~N.~Wang,
  Phys.\ Rev.\ Lett.\  {\bf 98}, 212301 (2007).

\bibitem{Armesto:2009zi}
  N.~Armesto, M.~Cacciari, T.~Hirano, J.~L.~Nagle and C.~A.~Salgado,
  J.\ Phys.\ G {\bf 37}, 025104 (2010).

\bibitem{alice}   K.~Aamodt {\it et al.} [ALICE Collaboration],
  JINST {\bf 3 }, S08002 (2008).

\bibitem{mult_paper} 
  K.~Aamodt {\it et al.} [ALICE Collaboration], Phys. Rev. Lett. {\bf 106}, 032301 (2011).

\bibitem{hijing}
  X.~N.~Wang and M.~Gyulassy,
  Phys.\ Rev.\  D {\bf 44}, 3501 (1991);

\bibitem{pythia6}
  T.~Sjostrand,
  Comput.\ Phys.\ Commun.\  {\bf 82}, 74 (1994).

\bibitem{perugia0}
  P.~Z.~Skands,
  Phys.\ Rev.\  D {\bf 82}, 074018 (2010).

\bibitem{geant3}
R.~Brun et al., 1985 GEANT3 User Guide, CERN Data Handling Division DD/EE/841 and
1994 CERN Program Library Long Write-up, W5013, GEANT Detector Description and Simulation Tool.

\bibitem{zyam}
  C.~Adler {\it et al.} [ STAR Collaboration ],
  Phys.\ Rev.\ Lett.\  {\bf 90}, 082302 (2003).

\bibitem{zyam_critisism}
  M.~Luzum,
  Phys.\ Lett.\  B {\bf 696}, 499 (2011).

\bibitem{zyam_crit2}
  T.~A.~Trainor,
  Phys.\ Rev.\  C {\bf 81}, 014905 (2010).

\bibitem{newflow}
  K.~Aamodt {\it et al.} [ALICE Collaboration],
  Phys.\ Rev.\ Lett.\  {\bf 107}, 032301 (2011).

\bibitem{ampt}
  Z.~W.~Lin, C.~M.~Ko, B.~A.~Li, B.~Zhang and S.~Pal,
  Phys.\ Rev.\  C {\bf 72}, 064901 (2005).

\bibitem{phojet}
  R.~Engel, J.~Ranft, S.~Roesler, Phys. Rev. D \textbf{52}, 1459 (1995).

\bibitem{renknew}
  T.~Renk, K.~.J.~Eskola,
  [arXiv:1106.1740 [hep-ph]].


\bibitem{iaastar2}
  B.~I.~Abelev {\it et al.} [ STAR Collaboration ],
  Phys.\ Rev.\  {\bf C80}, 064912 (2009).

\bibitem{jaa}
  A.~Adare {\it et al.} [PHENIX Collaboration], 
	Phys.\ Rev.\ C {\bf 78}, 014901 (2008).


\end{thebibliography}
\end{document}